**Local chemical order suppresses grain boundary migration under irradiation in CrCoNi**

Ian Geiger [1], Penghui Cao [2,3], Timothy J. Rupert [1,2,4,5,*]

[1] Materials and Manufacturing Technology, University of California, Irvine, CA 92697, USA

[2] Department of Materials Science and Engineering, University of California, Irvine, CA 92697, USA

[3] Department of Mechanical and Aerospace Engineering, University of California, Irvine, CA 92697, USA

[4] Hopkins Extreme Materials Institute, Johns Hopkins University, Baltimore, MD 21218, USA

[5] Department of Materials Science and Engineering, Johns Hopkins University, Baltimore, MD 21218, USA

* Corresponding author: tim.rupert@jhu.edu

## Abstract

Complex concentrated alloys with intrinsic chemical heterogeneity are promising candidates for nuclear applications, where local chemical order can strongly influence defect evolution under irradiation. Grain boundaries also contribute to radiation damage mitigation by serving as defect sinks, yet this interaction can alter interfacial structure, typically leading to destabilization and grain growth. This study investigates how chemical ordering influences grain boundary migration and stability during successive radiation events in CrCoNi. Using atomistic simulations, bicrystals were equilibrated to induce segregation-enhanced chemical order, followed by prolonged irradiation at 1100 K. Our results show that grain boundaries in random CrCoNi begin to migrate after only a few collision cascades, whereas those in the ordered alloy remain immobile until the chemical order is sufficiently disrupted. Single-cascade simulations reveal key mechanistic differences, where cascades near chemically ordered interfaces produce smaller damage volumes and reduced atomic displacement due to enhanced Frenkel pair combination within the cascade core. This limits both the residual defect population and the energetic driving force for interfacial rearrangement. Subsequent simulations of irradiated interfaces show that interstitial absorption induces a structural transition that modifies the segregation morphology at and near the grain boundary, demonstrating a dynamic coupling between ordering stability and defect evolution. These findings offer new insights into the role of local chemical order on defect-interface interactions under extreme conditions and highlight pathways for designing radiation-tolerant materials for next-generation nuclear systems.

## 1. Introduction

Grain boundaries are critical structural features that govern the mechanical properties and irradiation tolerance of polycrystalline materials. A widely used strategy for enhancing strength involves stabilizing a nanocrystalline grain structure (grain sizes below 100 nm), where the high volume fraction of grain boundaries creates a dense network of obstacles to plastic deformation.[1,2] Nanostructured materials have also attracted growing interest in nuclear applications due to their abundance of interfaces, which can serve as efficient sinks for irradiation-induced defects.[3–5] However, the elevated temperatures and intense particle fluxes characteristic of current and next-generation nuclear reactors make nanocrystalline materials highly susceptible to microstructural evolution, including grain growth,[6–8] which can degrade structural integrity and reduce defect sink efficiency. Ensuring reliable performance requires a fundamental understanding of the mechanisms that mediate defect formation and stability in such extreme environments.

During irradiation, high-energy particles transfer kinetic energy to primary knock-on atoms (PKAs), displacing them from their parent lattice sites and initiating a displacement cascade. These cascades generate localized thermal spikes and produce a dense population of point defects (i.e., interstitials and vacancies) that can aggregate and drive changes in material properties.[9,10] Experimental studies of polycrystalline metals and ceramics show that average grain size increases with irradiation dose, often following a power-law dependence.[11–14] Grain boundaries near thermal spikes are particularly prone to migration,[11] even at overall sample temperatures where conventional thermally-activated boundary motion would be negligible.[12,15] Kaoumi et al.[11] proposed a model of grain growth based on the interaction of radiation-induced thermal spikes with grain boundaries, linking boundary mobility to both irradiation parameters and intrinsic material properties such as cohesive energy and thermal conductivity. In this model, atomic jumps within the thermal spike region, biased by boundary curvature, drive migration to reduce interfacial area. Molecular dynamics (MD) simulations support this mechanism, showing that cascades overlapping grain boundaries can trigger migration.[16,17] Additional simulations of planar boundaries in nanocrystalline Cu further reveal that interstitial loading and defect cluster interactions can restructure interfaces and initiate boundary migration, followed by curvature-driven growth.[18]

Since irradiation-induced grain growth is fundamentally diffusion-controlled, materials with greater diffusion barriers, defect formation energies, or Frenkel pair recombination rates (enhanced, for example, by prolonged thermal spike durations in low thermal conductivity systems) are expected to exhibit reduced boundary mobility and improved coarsening resistance. Complex concentrated alloys (CCAs), composed of multiple principal elements in near-equimolar ratios, provide a compelling platform for investigating these behaviors.[19] Unlike conventional metals and alloys, CCAs have pronounced atomic-scale chemical complexity, producing a rugged potential energy landscape that frustrates the stability and transport of point defects,[20,21] dislocations,[22–24] and grain boundaries.[25–28] These effects are further intensified by the emergence of nanoscale chemical domains, or local chemical ordering (LCO), arising from short-range enthalpic correlations between atomic species.[29,30] LCO imposes site-specific variations in migration energies and jump frequencies, leading to anisotropic and spatially heterogeneous diffusion kinetics that slow point defect migration.[19,31,32] In CrCoNi, restricted migration pathways, combined with increased point defect formation energies, enhance Frenkel pair recombination during thermalization by minimizing the diffusivity disparity between vacancies and interstitials and promoting local defect annihilation.[33,34] Similar improvements in irradiation resistance have been observed in other Ni-containing CCAs, where higher recombination rates, suppressed defect clustering,[35,36], and lower thermal conductivity (which extends the thermal spike duration and enhances recombination efficiency)[37,38] all contribute to superior radiation tolerance.

While the diffusion behavior of point defects in CCAs across different states of chemical ordering has been relatively well studied in several alloy systems,[20,31,39,40] the effects of irradiation-driven defect evolution on grain boundary stability remain unclear. Enhanced chemical binding in low-energy LCO states may raise the activation barrier for atomic rearrangements during thermal spikes, thereby suppressing atomic jumps across the boundary and limiting interfacial migration. Such a mechanism was proposed by Zhou et al.[41] to explain the increased grain growth rates for $Al_x$CoCrFeNi CCAs with higher Al content. Beyond this effect, LCO may also inhibit grain boundary motion by reducing the driving force for migration. As a boundary moves, it disrupts low-energy chemical configurations that must be reestablished in a new crystallographic orientation, imposing an energetic penalty that slows migration.[42] This behavior is particularly relevant in CCAs, where segregation and LCO are often enhanced at grain boundaries.[43,44] For example, in CrCoNi and NbMoTaW, grain boundary segregation and vicinal structural disorder have been shown to amplify anisotropic LCO near interfaces.[45] A further complication arises from the dynamic chemical reordering that occurs during the ballistic and thermalization stages of radiation cascades. Depending on the initial configuration, this reordering may either reinforce or degrade preexisting chemical order with concomitant changes in defect behavior.[46,47] For example, Zhang et al.[33] reported that CrCoNi with initial LCO exhibits smaller average defect cluster sizes and lower number densities than a random solid solution counterpart at low damage doses (<0.6 displacements per atom or dpa). However, these differences diminish at higher irradiation levels, suggesting that cascade-driven reordering eventually erodes the stabilizing influence of LCO. A comparable, transient phenomenon is likely to influence grain boundary behavior as boundaries are progressively driven out of equilibrium under sustained irradiation.

In this work, atomistic simulations are used to investigate the influence of LCO on grain boundary evolution and migration at a Σ5(310) symmetric tilt boundary in CrCoNi under irradiation. CrCoNi has been widely studied for its ordering tendencies,[48,49] with recent experiments showing that LCO can persist over a wide temperature range.[50] Bicrystalline models were equilibrated with hybrid Monte Carlo (MC)/MD simulations to generate LCO (*Segregated*), then annealed and subjected to consecutive radiation damage events. Comparisons of grain boundary behavior were made to chemically disordered bicrystals (*Disordered*). Displacement cascades were found to disrupt LCO in Segregated samples but slightly enhance it in Disordered systems, with chemical order parameters converging after ~300-400 cascades. In the early cascade regime (<100 cascades), Segregated boundaries remain largely immobile, whereas Disordered boundaries exhibit immediate migration and structural fluctuations. Single PKA events near the boundary further reveal that Disordered systems are more susceptible to cascade-induced distortion, developing pronounced boundary curvature and vacancy-rich cavities that drive atomic migration. Although thermal spike core temperatures are comparable in both cases, Disordered systems exhibit higher atomic displacement due to an increased tendency for atomic jumps. This amplifies boundary fluctuations and migration, and offers a pathway to limit grain growth under irradiation by engineering LCO into the microstructure.

## 2. Computational Methods

Atomic interactions in the CrCoNi system were modeled using an embedded atom method (EAM) potential developed by Li et al.[51] This potential has been widely employed to investigate the effects of configurational state on defect stability and transport behavior.[26,33,34,52] Prior studies have shown that LCO can evolve and intensify at extended defects such as dislocations[53] and grain boundaries[45] in CrCoNi, making it an ideal candidate for exploring the influence of LCO on grain boundary stability. A known limitation of the Li et al. potential is that it does not fully reproduce all pairwise chemical correlations relative to DFT. For example, while the EAM potential predicts weaker Ni-Cr ordering than DFT-based studies,[54,55] both methods consistently predict Co-Cr ordering and Cr-Cr repulsion. This difference has been attributed to limitations in the DFT data used during potential fitting, including the absence of explicit magnetic effects.[51] DFT-informed machine-learning interatomic potentials have since been developed to improve chemical accuracy in CrCoNi.[56,57] However, the substantially higher computational cost of these potentials limits accessible system sizes and equilibration times, making them impractical for modeling the large bicrystal geometries and repeated cascade simulations required in this work. Nonetheless, exploratory simulations performed using the neural network potential on smaller systems over fewer equilibration steps show the onset of similar near-boundary segregation trends,[45] providing confirmation of the spatial localization of LCO at grain boundaries in CrCoNi.

A $\Sigma 5$ (310) symmetric tilt boundary was chosen for this study as a representative high-angle boundary, having been previously characterized in the context of radiation-induced grain boundary migration in pure nanocrystalline metals.[59] Atomistic simulations were performed using the Large-scale Atomic/Molecular Massively Parallel Simulator (LAMMPS) software package.[60] The equilibrium boundary structure was obtained via an iterative sampling method.[61] Simulation cells contained 238,800 atoms with dimensions of 11.4 nm × 22.8 nm × 10.8 nm and periodic boundary conditions applied in all directions. Initial configurations were constructed from pure

Ni, followed by random substitution of Ni atoms with Co and Cr to achieve an equimolar composition.

To generate atomic configurations with grain boundary segregation and LCO, an initially random solid solution was annealed at 300 K for 50 ps using MD with 1 fs timestep, followed by hybrid MC/MD simulations until chemical and structural convergence were achieved ($\sim 4 \times 10^6$ MD steps). Every 20 MD steps, one MC step was performed based on the Metropolis criterion, with atom swaps attempted for one-quarter of the atoms. MC/MD simulations were performed in the NPT ensemble at zero pressure using the variance-constrained semi-grand canonical ensemble (VC-SGC), which overcomes the limitations of the canonical and semi-grand canonical ensembles in modeling precipitation and phase segregation. Chemical potential differences of $\Delta\mu_{Ni\text{-}Co} = 0.021$ eV and $\Delta\mu_{Ni\text{-}Cr} = -0.031$[51] were used to maintain equiatomic compositions, with the variance parameter $\kappa = 10^3$. This process was repeated to yield four thermodynamically equivalent bicrystalline samples with local chemical ordering. An additional four specimens were generated by extending each of the original MC/MD runs by $5\times10^4$ MC/MD steps, producing eight equivalent, but configurationally distinct, samples (*Segregated*, Fig. 1(a) and (c)). Eight random solid solutions (*Disordered*, Fig. 1(b) and (d)) were also produced for comparison. It should be noted that LCO evolves dynamically during irradiation in both systems; therefore, the terms Segregated and Disordered refer only to the initial chemical configurations and do not represent static chemical states throughout the simulation. Each sample was subsequently annealed at 1100 K to emulate thermal conditions relevant to next-generation reactors[62] and to provide sufficient thermal activation for grain boundary motion on MD timescales.

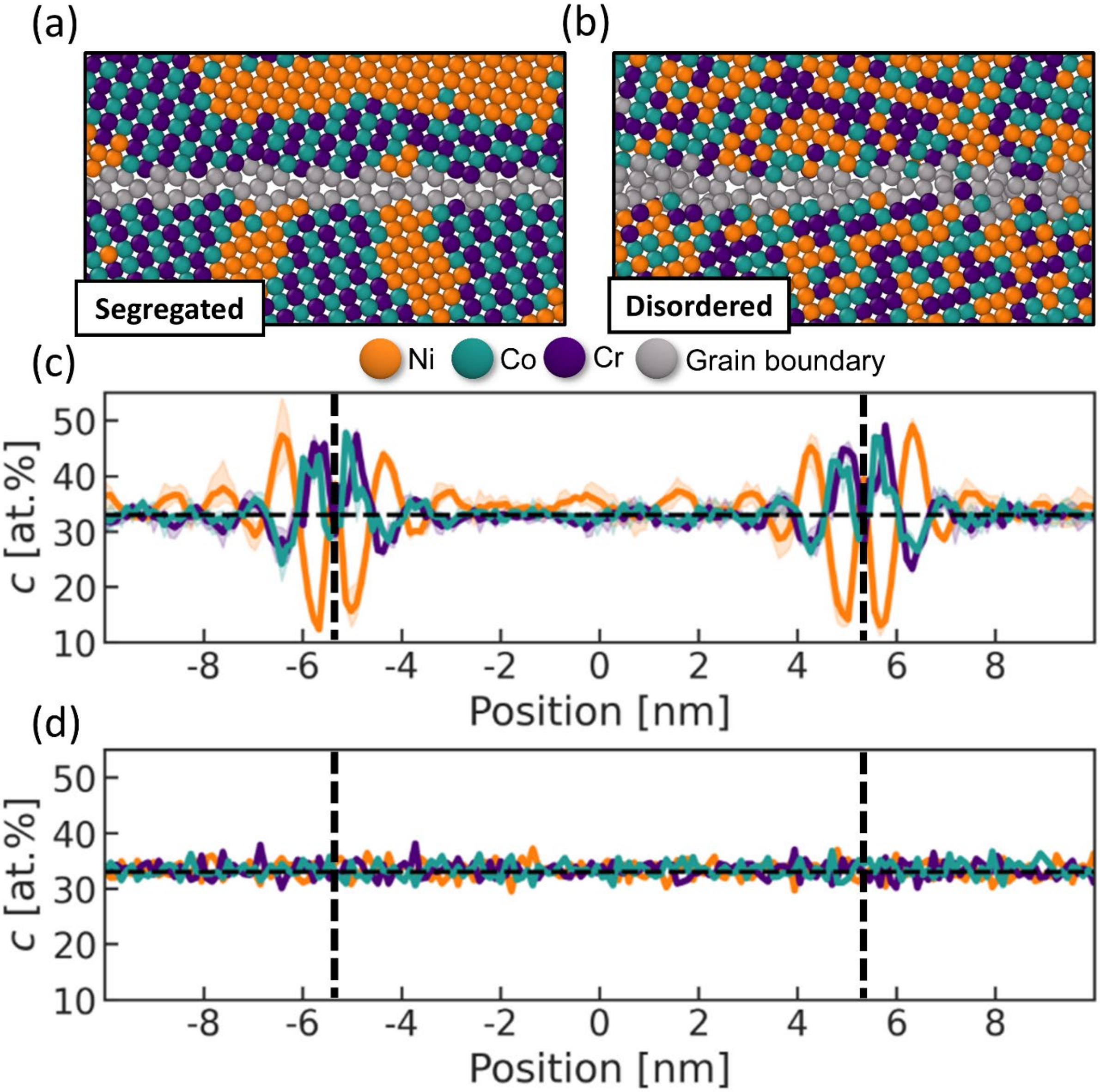

**FIG 1. Atomistic data for grain boundary composition. In (a) and (b), snapshots of atomic configurations are shown for Segregated and Disordered chemical states, respectively. Grain boundary atoms are colored gray to highlight the composition in the adjacent regions. In (c) and (d), elemental concentrations for Segregated and Disordered conditions, respectively, are shown where dashed lines represent the centers of each grain boundary.**

To achieve an adequate damage dose and capture defect evolution under irradiation, a successive displacement cascade algorithm was used.[63] The Ziegler-Biersack-Littmark repulsive potential was smoothly joined with the EAM potential to accurately model high-energy atomic collisions,[64,65] while an adaptive time step algorithm was used to limit atomic displacements to 0.02 Å per step to ensure numerical stability. For each cascade, a randomly selected atom was assigned as the PKA with a kinetic energy of 5 keV and a random incident direction. The simulation cell was translated to place the PKA at the cell center, minimizing interactions with the boundary regions, and 45,000 adaptive timesteps were performed. Cascades were simulated in the microcanonical (NVE) ensemble, with a Nosé-Hoover temperature-rescaling thermostat applied at all boundaries (one lattice constant thick) to absorb excess energy and maintain the system near 1100 K. After each cascade and subsequent relaxation, the simulation cell was shifted back to its original position to enable consistent defect tracking across cascades. This procedure was repeated until 400 cascades were completed, a sufficient number to capture the impact of LCO on grain boundary evolution. Defects and grain boundary structure were analyzed using Polyhedral Template Matching (PTM)[66] and the Wigner-Seitz cell method in OVITO.[67] Structural identification employed an RMSD cutoff of 0.1 in the PTM analysis, and Wigner-Seitz defect analysis was performed relative to the defect-free pre-irradiation configuration.

## 3. Results and Discussion

### 3.1 Coupled evolution of grain boundaries and local chemical ordering under irradiation

The effect of irradiation-induced atomic mixing on LCO during consecutive cascades was investigated first to contextualize grain boundary behavior. Warren-Cowley order parameters[68] were calculated for all face-centered cubic (FCC) atoms and are shown for Ni-Ni and Co-Cr pairs in Fig. 2, with solid and dashed lines denoting first- and second-neighbor interactions, respectively. This pairwise order parameter is defined as

$$\alpha_n^{ij} = 1 - \frac{p_n^{j,i}}{c_j}, \quad (1)$$

where $\alpha_n^{ij}$ quantifies the chemical correlation between a central atom of type $i$ and atoms of type $j$ in the $n$-th neighbor shell. $p_n^{j,i}$ is the local concentration of species $j$ in the $n$-th shell surrounding atoms of type $i$, and $c_j$ is the global concentration of species $j$. Negative values of $\alpha_n^{ij}$ indicate a greater probability of finding $j$ atoms near $i$, positive values indicate avoidance, and values near zero correspond to random mixing. Accordingly, the near zero values observed for all atomic pairs in the Disordered configurations indicate the absence of LCO.

With increasing displacement cascades, the parameters for each atomic pair in Disordered samples become negative (attractive) due to the formation of energetically favorable atomic configurations. The slope stabilizes near 200 cascades, indicating a balance between irradiation-induced mixing and thermodynamically-driven segregation. In contrast, the initially strong LCO in the Segregated samples is progressively destroyed, decreasing by more than half for both pairs by 100 cascades. Consequently, the effect of LCO on grain boundary stability is expected to be greatest during the early stages of damage accumulation (fewer than 100 cascades in this study). A weaker overall trend in second nearest-neighbor interactions (dashed lines in Fig. 2) indicates that reordering is primarily occurring in the first coordination shell at this irradiation temperature. After 300-400 cascades, the order parameters converge.

The evolution of LCO in Fig. 2 is consistent with a model proposed by Li et al.[69] in which LCO disrupted by bombardment partially re-forms during thermalization through radiation-enhanced diffusion. Similar interplay between irradiation-induced atomic mixing and LCO reconstruction has also been invoked to explain the increased area fraction of LCO observed after high-temperature irradiation in NiCoFeCrMn.[47] The resulting plateau in LCO parameters therefore represents a steady-state balanced by the competition between cascade-induced atomic mixing, which destroys order, and thermally activated chemical relaxation between cascades, which partially restores favorable local configurations at 1100 K. Under lower flux conditions, longer intervals between cascades would allow greater diffusion-mediated recovery, and the steady-state LCO would be expected to shift toward the equilibrium ordering level at the irradiation temperature. The magnitude of this shift is limited by the reduced thermodynamic stability of LCO at 1100 K relative to lower temperatures.

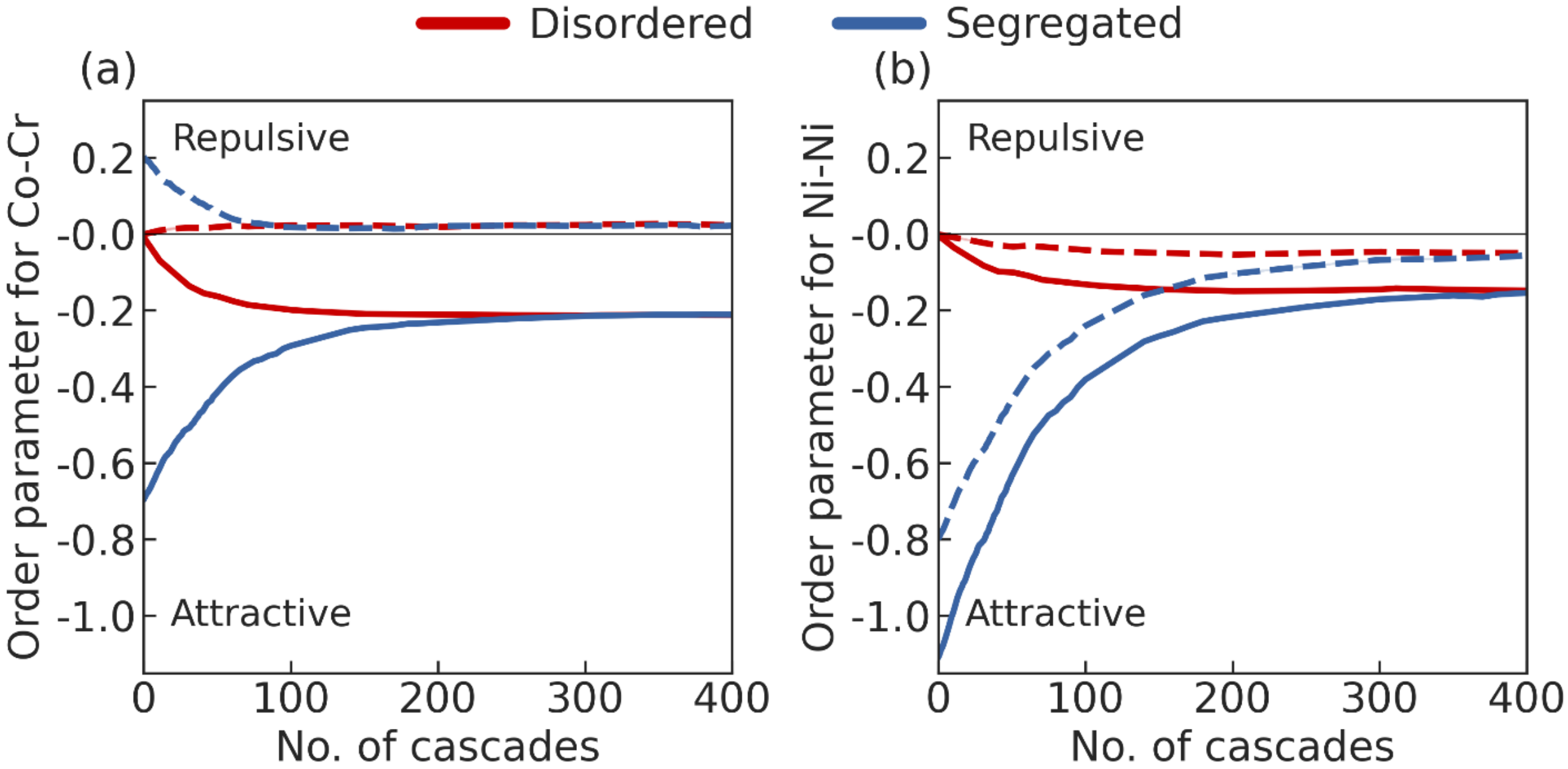

**FIG 2. Evolution of chemical order parameters under irradiation. In each plot, the solid and dashed lines represent the first- and second-nearest neighbor Warren-Cowley order parameters, respectively, for each atomic pair.**

Defect evolution in the two chemical states was examined next with emphasis on grain boundary evolution and migration. Figs. 3(a-d) present atomic snapshots of defects and boundary atoms for a representative Disordered sample, while Figs. 3(e-f) show those for a Segregated sample. Snapshots are taken from the initial state and after 50, 100, and 200 cascades, when LCO differences are most pronounced. Initially, both sets of grain boundaries are aligned at identical positions (dashed blue lines in Figs. 3(a) and (e)). Disordered boundaries exhibit a rough morphology from local roughening transitions driven by compositional fluctuations,[70] whereas Segregated boundaries remain atomically flat. After 50 cascades, Disordered boundaries migrate rapidly toward the cell center, accompanied by the buildup of vacancies and interstitials, as well as larger vacancy clusters that aggregate into stacking fault tetrahedra (SFT). Migration in the Disordered cell continues until the interfaces coalesce after ~200 cascades here (Fig. 3(d)), though not all Disordered boundaries coalesce by the end of the simulation run. In contrast, despite comparable defect accumulation within the grains, Segregated boundaries remain pinned near their initial positions through at least 100 cascades. Even at 200 cascades (Fig. 3(h)), they display only limited structural fluctuations and boundary displacement.

Although irradiation-induced defects evolve with each consecutive cascade, their density appears to saturate due to competing processes such as grain boundary-biased flux, vacancy-interstitial recombination, and defect clustering. To understand how these processes contribute to interfacial migration, it is useful to track vacancy and interstitial populations through Frenkel pair analysis, since excess point defect concentrations can generate fluxes that drive boundary motion and affect stability.[59,71,72] A common approach for this analysis utilizes the Wigner-Seitz cell method, which compares atomic site occupancies in a defect-free reference to those in the displaced configuration after irradiation. However, once boundaries migrate, this method becomes unreliable because atoms reoriented into new crystalline orientations are incorrectly classified as point defects. For this reason, quantitative Frenkel pair analysis is only applied during early cascade simulations before large-scale migration occurs, as discussed later.

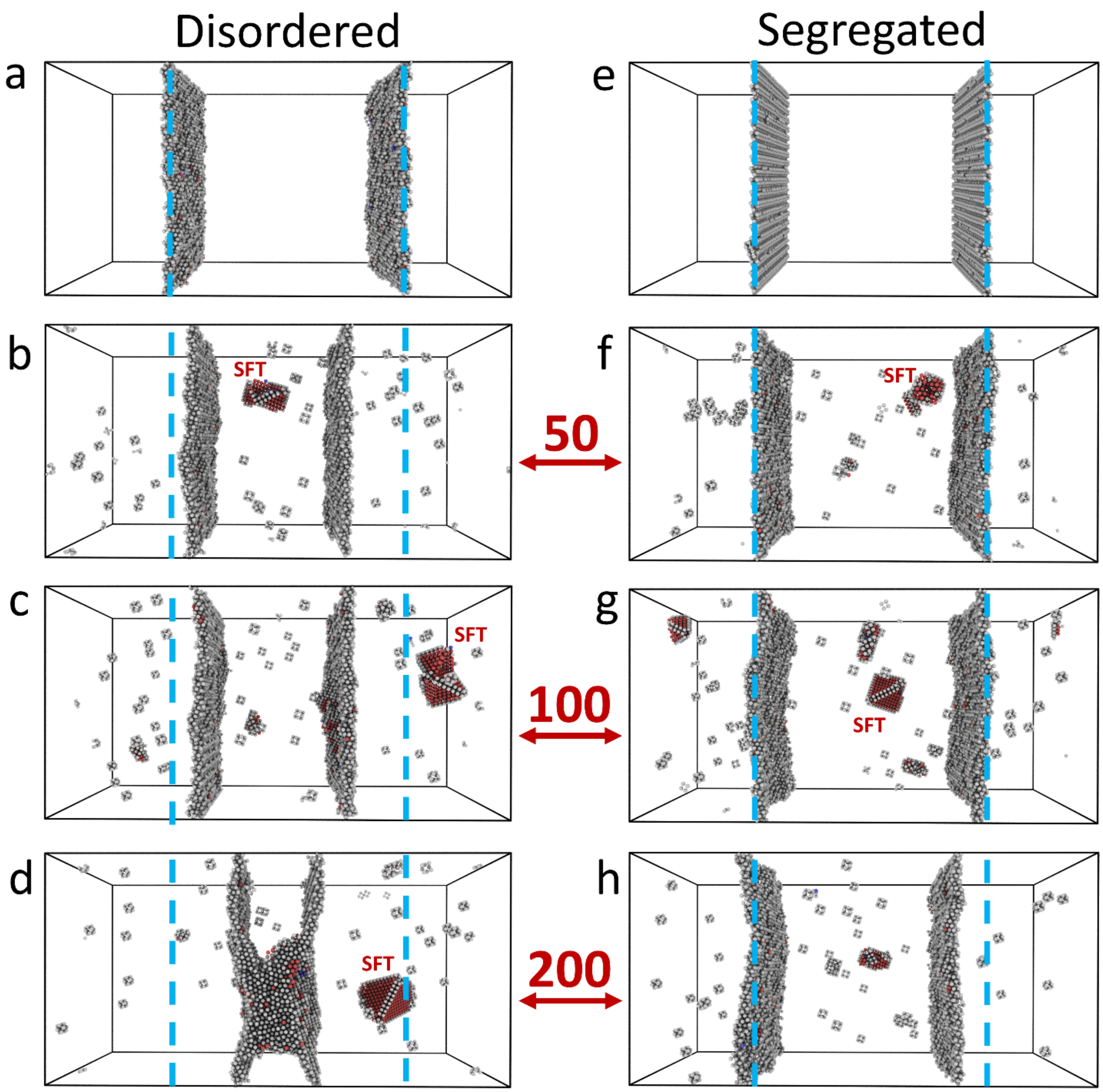

**FIG 3. Irradiation response of representative Disordered (a-d) and Segregated (e-h) grain boundaries during the first 200 cascades. Snapshots show non-FCC atoms in the initial configurations (a, e) and after 50, 100, and 200 cascades at 1100 K. FCC atoms are omitted for clarity. Atoms are colored by structure type using PTM, where gray denotes structurally disordered atoms and red denotes hexagonal close-packed atoms. The dashed blue line indicates the initial interface position.**

During irradiation, the mean grain boundary positions, $\bar{h}$, were recorded and plotted by cascade in Fig. 4(a) and (b) for Disordered and Segregated boundaries, respectively. Each color represents a unique simulation run. The boundary positions were adjusted so that the mean boundary positions are visualized relative to their initial position, regardless of which specific grain boundary is being measured. Within the first 100 cascades, $\bar{h}$ deviates significantly from its initial position for Disordered boundaries but remains largely dormant for Segregated ones. Coalescence occurs at ~100 cascades for one set of Disordered grain boundaries and near 200 cascades for another. Dashed lines in Fig. 4(a) show linear fits of the grain boundary migration trajectory for these boundaries, so that their rate of migration can be compared to other samples. Other boundaries in Fig. 4(a) exhibit classical random walk behavior throughout the simulation, reflecting volatility caused by the combined influence of collision cascade, thermal diffusion, and defect-interface interactions. Since each PKA and its velocity vector are randomly selected, the lack of an applied driving force leads to stochastic migration. The boundaries behave like sails absorbing random "gusts" of energy from cascades, producing erratic displacements without a consistent direction. In some cases, consecutive displacement cascades in aligned directions can collectively bias migration, eventually leading to coalescence. In contrast, significant deviations in $\bar{h}$ for Segregated boundaries emerge only after ~100 cascades and exhibit much milder slopes. At higher doses (between 100-400 cascades), their motion becomes intermittent, resembling the abrupt jumps seen in Disordered boundaries, though the interfaces never reach close enough proximity to interact.

Given the elevated simulation temperature, a natural question is whether the kinetic activation of the grain boundary is purely a thermal annealing effect. To clarify this, additional MD simulations were conducted at 1100 K without irradiation. As discussed in Supplementary

Note 1, only very minimal grain boundary migration occurs over the same timescales without irradiation. This indicates that grain boundary evolution differs in both magnitude and mechanism, confirming that annealing alone cannot induce the same level of migration.

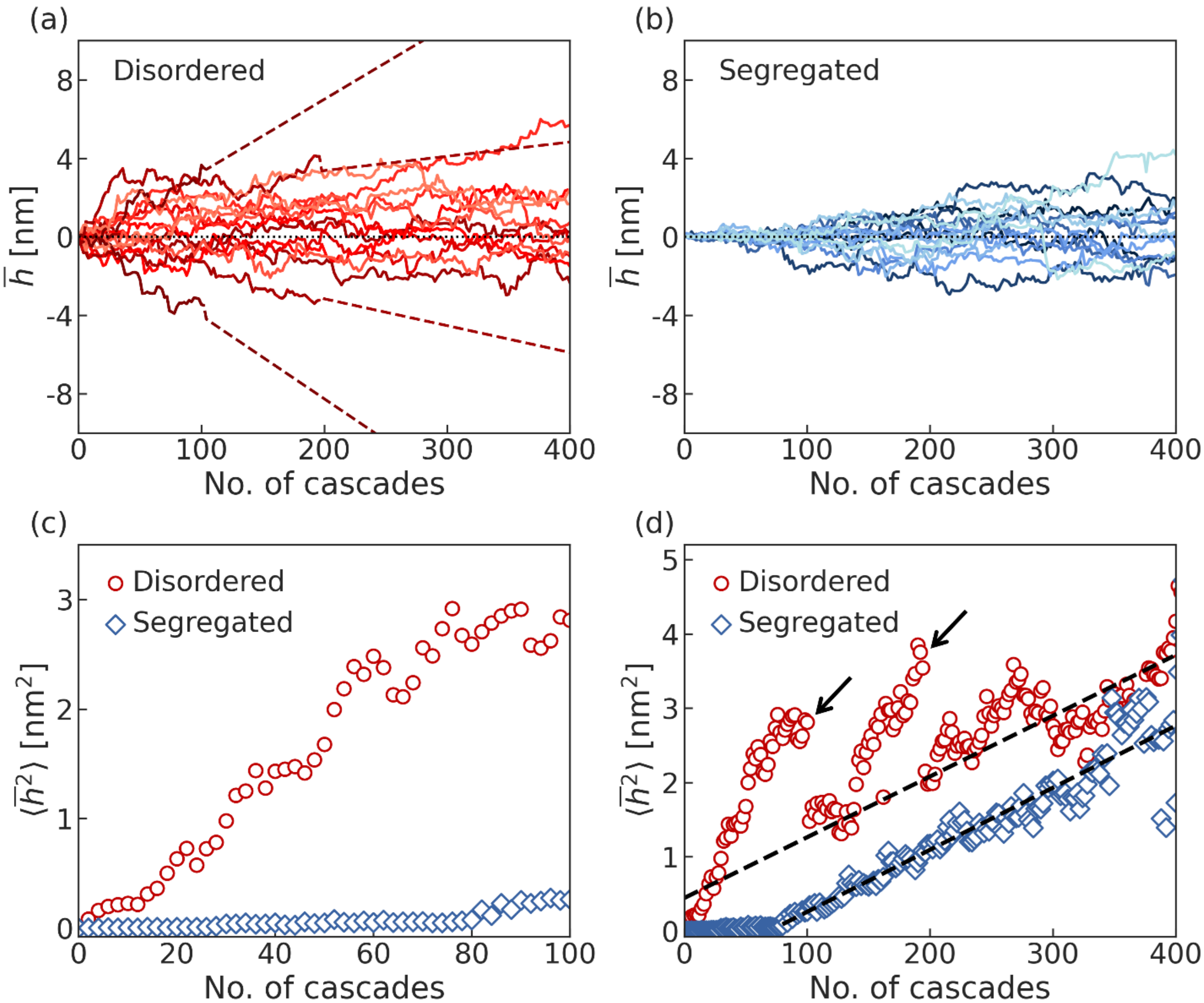

**FIG 4. Grain boundary migration statistics. (a-b) Average grain boundary displacement, $\bar{h}$, for each unique simulation in (a) Disordered and (b) Segregated boundary sets. Line color indicates the specific grain boundary pair in each bicrystal (8 total). (c) Variance in grain boundary position, $\langle \bar{h}^2 \rangle$, calculated from the data in (a-b) over the first 100 cascades. (d) $\langle \bar{h}^2 \rangle$ for all Disordered and Segregated boundaries over 400 cascades. Arrows mark peaks that precede sharp drops associated with boundary coalescence. Dashed lines show linear fits applied after a clear increase in $\langle \bar{h}^2 \rangle$; coalesced Disordered boundaries were excluded from the fit in (d).**

Figs. 4(c) and (d) shows the temporal evolution of the variance in grain boundary position, $\langle \bar{h}^2 \rangle$, for the simulation sets presented in Figs. 4(a) and (b). This metric is commonly used to assess grain boundary mobility.[73] The comparison reveals clear statistical differences in migration behavior between initially chemically disordered (Disordered) and initially chemically ordered (Segregated) boundaries. In the Disordered samples, $\langle \bar{h}^2 \rangle$ increases rapidly within the first ~10 cascades, while in the Segregated samples it only begins to rise after 80 cascades (Fig. 4(c)). Even after this onset, the spread in $\bar{h}$ for Segregated boundaries (Fig. 4(b)) grows gradually, indicating sustained migration suppression. Extending the analysis to 400 cascades (Fig. 4(d)) shows that variance in Segregated boundary displacement increases monotonically after the delayed onset, whereas Disordered boundaries display more erratic growth due to pervasive chemical disorder and boundary instability. Arrows in Fig. 4(d) mark sharp drops in $\langle \bar{h}^2 \rangle$ caused by boundary coalescence in a given simulation run, which reduces the sample size and thus artificially lowers the apparent variance. Linear fits were applied to each dataset, excluding coalesced boundaries to avoid bias. For the Segregated set, the fit begins at cascade 80, after the onset of large-scale migration. Both fits yield similar slopes (0.008 $nm^2$/cascade), indicating that while the Disordered boundaries become mobile quickly, Segregated boundaries undergo a delayed, gradual release from their initially pinned states before fluctuating freely once residual LCO is overcome.

The degree of initial chemical order clearly plays a key role in the grain boundary suppression observed in Figs. 3 and 4. While the Disordered and Segregated configurations represent two limiting cases, irradiation progressively drives the system toward intermediate states with reduced LCO and smaller ordered domains. LCO in the Segregated samples gradually converges to the level present in the Disordered samples. As this occurs, grain boundary fluctuations increase with cascade accumulation and eventually approach similar rates in both systems. These results indicate that stabilization scales with the magnitude of LCO during irradiation rather than reflecting a binary ordered versus disordered response.

Although grain boundary migration is significantly delayed in the Segregated samples, under experimental conditions with longer relaxation times and lower dose rates, atoms displaced from ordered structures would have greater opportunity to diffuse back toward energetically favorable configurations. Such recovery of LCO could further suppress interface migration by reforming ordered regions near the boundary. In our simulations, cascades were introduced every 35-40 ps, set by the adaptive time-step, which is sufficient to capture both ballistic and thermalization phases as the system returns to steady state at 1100 K. This interval, chosen for computational efficiency, is several orders of magnitude shorter than the time scales accessible in experiments. Recent MD studies of Ni and Ni-based alloys have shown good agreement with experimental observations of defect structures,[63,74] suggesting that the present simulations accurately capture the initial grain boundary response. However, the long-term interplay among defect production, irradiation-induced mixing, and diffusion-enabled LCO recovery remains beyond MD-accessible time scales. It is therefore reasonable to expect that at lower effective fluxes or with longer intervals between cascades, resistance to boundary migration could exceed that observed here. For reference, the simulations reached only ~0.1 dpa based on the NRT approximation,[75] whereas Zhang et al. reported pronounced differences in average size of defect clusters between random and LCO CrCoNi samples up to 0.6 dpa,[33] likely due to longer relaxation times between successive irradiation events that enabled more effective LCO recovery.

### 3.2. Atomistic mechanism for suppressed grain boundary migration in Segregated CrCoNi

Previous studies have shown that collision cascades near grain boundaries generate thermal spikes and excess point defects, which directly influence boundary migration.[11,16,76] However, the markedly different migration behaviors observed in the Segregated and Disordered samples suggests that these mechanisms are significantly altered by the presence of LCO. To elucidate these mechanisms, single PKA simulations were performed. In each sample, a 5 keV PKA was initiated 5 Å from the grain boundary plane, with momentum directed normal to the interface. Fig. 5 shows atomic snapshots at 2 ps, 20 ps, and 30 ps for atoms within the grain boundary and cascade regions, colored by their position along the normal direction. The mean displacement ($\Delta\bar{h}$) quantifies average atomic displacement relative to the initial interface position, while the standard deviation (σ) measures the extent of interfacial fluctuations. Additional simulations for each sample type, with randomly selected PKA locations also positioned 5 Å from the boundary, are presented in Supplementary Note 2.

A typical cascade proceeds through a ballistic phase (~0.1-0.2 ps), during which atoms are rapidly displaced, followed by a thermal spike where the cascade core behaves as a transient liquid before quickly thermalizing with the surrounding lattice.[77] At 2 ps (Fig. 5, top row), the thermal spike forms a roughly cylindrical displacement field beneath each boundary with $\Delta\bar{h}$ reaching -7.7 Å and -3.1 Å in the Disordered and Segregated boundaries, respectively. By 20 ps, the Disordered boundary still shows substantial structural disorder, whereas the Segregated boundary has largely recovered, indicating more localized damage and greater efficiency in Frenkel pair recombination in the presence of LCO. Wigner-Seitz analysis reveals deeper and more numerous residual vacancies and interstitials in the Disordered sample. In both cases, a vacancy concentration gradient develops and increases with distance from the boundary as mobile interstitials are preferentially absorbed by the interface.[5,78] In the Disordered sample, the higher density of excess vacancies promotes grain boundary migration by creating a cavity of empty sites that facilitate atomic rearrangement and generate a concave distortion in the initially flat interface. After 30 ps, this results in pronounced waviness in the Disordered sample, while the Segregated sample shows only a slight bulge. The difference is quantified by σ at 30 ps, which is 5.0 Å and 2.2 Å for Disordered and Segregated samples, respectively. The total boundary displacement is also roughly five times greater in the Disordered case. Notably, single PKAs initiated within the grains do not affect the boundary structure.

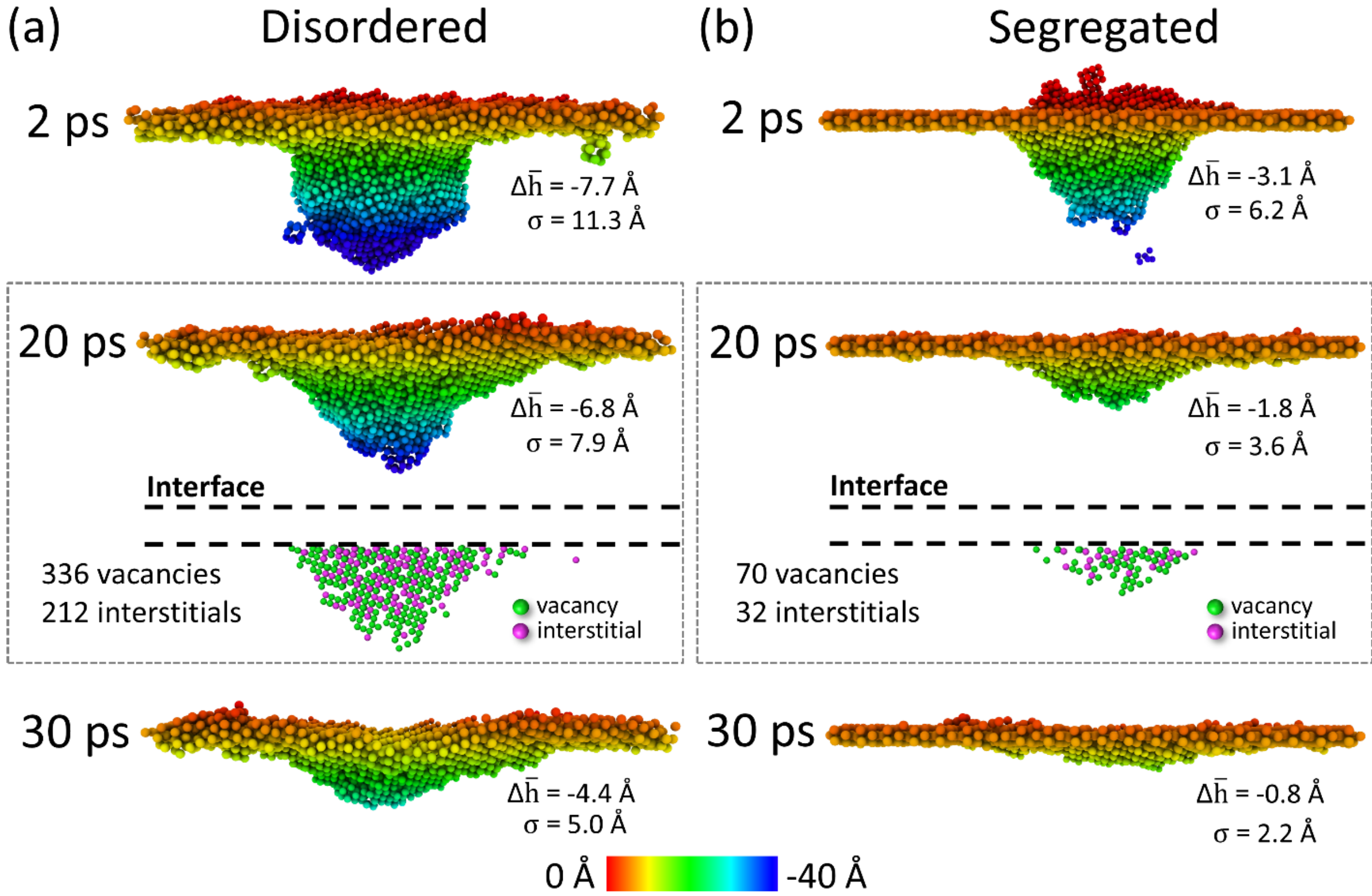

**FIG 5. Evolution of single-PKA collision cascades during the thermal spike phase for (a) Disordered and (b) Segregated boundaries. Non-FCC atoms are shown and colored by depth after 2 ps, 20 ps, and 30 ps. At 20 ps, Wigner-Seitz cell analysis highlights point defect distributions and vacancy gradients. The mean boundary displacement from the initial position, $\Delta\bar{h}$, and interfacial fluctuations, σ, are reported at each timestep to quantify migration and fluctuation behavior.**

The above findings demonstrated that the varying quantities of vacancy and interstitial between the Disordered and Segregated states lead to significant changes in the grain boundary morphology. Since the diffusion and recombination of these defects is related to the transient elevated temperatures during thermalization, the temperature and mean-squared displacement (MSD) of atoms in the cascade core were tracked for 100 ps during the single-PKA simulations and are presented in Fig. 6. The cascade core was initially defined as a 2.5 nm radius sphere centered on the center of mass of atoms with kinetic energy >1.2 eV at 0.1 ps. However, all atoms within this volume, regardless of energy, were included in subsequent calculations. Core temperature was calculated via $K = 3Nk_bT$, where $K$ is the total kinetic energy of $N$ atoms within the volume, $k_b$ is the Boltzmann constant, and $T$ is the temperature. This region remained fixed during the simulation for temperature calculations, with atoms moving in and out. For MSD, the original atoms in the cascade core volume were tracked throughout the simulation. Cascades from five randomly selected PKAs were analyzed with similar results, though Fig. 6 presents one representative dataset for clarity. Fig. 6(a) shows that initially high core temperatures stabilize over time to 1100 K due to thermal insulation at the simulation cell borders. In NiCoCrFe, enhanced Frenkel pair recombination within the cascade core of a LCO structure was attributed to a prolonged thermal spike due to reduced thermal conductivity.[37] However, in CrCoNi here, LCO does not affect the cascade core temperature since both curves in Fig. 6(a) decrease at the same rate, meaning this is not the primary factor underlying the enhanced damage recovery observed in Fig. 5. Instead, MSD trends in Fig. 6(b) show that the average displacement of atoms initially in the cascade region increases more quickly for Disordered samples compared to Segregated samples. The MSD decays exponentially (red and black dashed lines in Fig. 6(b)), with energy dissipation leading to reduced motion over time. The decay rate constant, $\lambda$, is 0.101 for the Disordered sample and 0.158 for the Segregated sample (a 56% increase), leading to atoms in the Disordered state migrating nearly twice the distance of those in the Segregated state on average.

The faster decay in the Segregated sample can be attributed to the presence of LCO, which impedes diffusion of both interstitials and vacancies, consistent with recent simulation studies.[20,21] LCO also reduces mobility disparities between point defects in CrCoNi, increasing recombination rates and limiting defect accumulation near the grain boundary.[33] Consequently, fewer interstitials migrate to the grain boundary and less vacancies are available to form a cavity, thereby reducing the driving force for atomic migration. As a result, MSD decays faster, and the relaxation time to steady-state drops from 30 ps in the Disordered sample to 20 ps in the ordered one. As LCO progressively degrades with continued irradiation, the divergence in MSD is expected to narrow, eventually converging to a common decay rate.

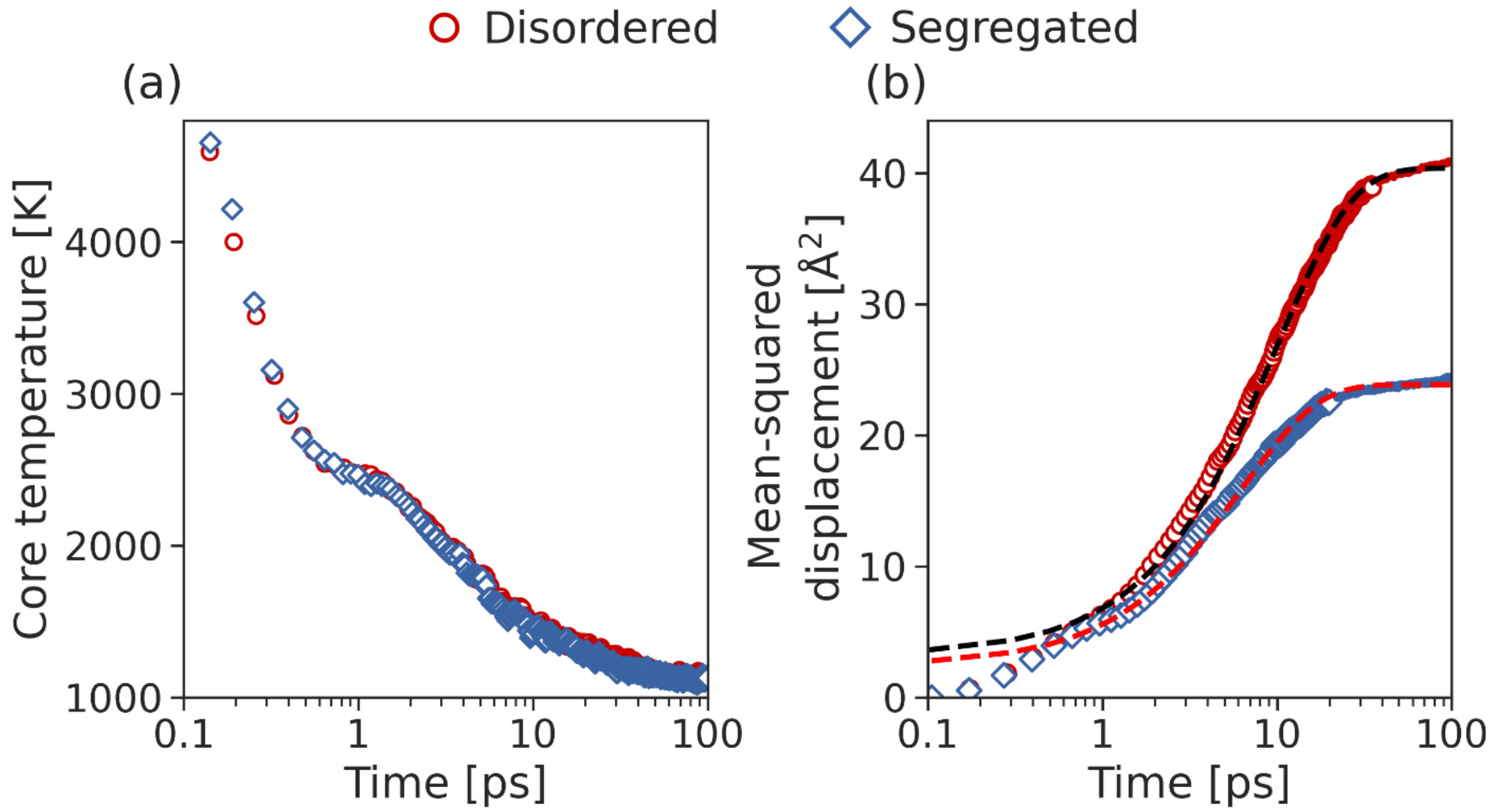

**FIG 6. Time-evolution of the (a) cascade core temperature and (b) mean-squared displacement of atoms within the initial core region. In (b), black and red dashed lines represent the exponential decay curve for the Disordered and Segregated states, respectively.**

The previous evidence suggests that large-scale grain boundary migration is preceded by localized atomic restructuring, driven by the accumulation of excess point defects from collision cascades. These restructuring events, captured by grain boundary fluctuations ($\sigma$ in Fig. 5), increase both the boundary area and energy. Over time, the boundary relaxes along its curvature to lower energy, consistent with thermal spike models,[11] leading to coordinated atomic displacements. To assess how this process relates to the delayed boundary migration observed in the prior consecutive cascade simulations, $\sigma$ was evaluated after each cascade and results binned in increments of 10 cascades in Fig. 7. Disordered boundaries consistently display higher $\sigma$ values and broader distributions than Segregated boundaries from the outset, with little evolution over time. This indicates greater structural fluctuations in Disordered samples, which can promote atomic rearrangements that enable migration to reduce curvature. The initial disparity in median values likely arises from the higher structural order present in the Segregated samples, an effect that diminishes as LCO and segregation progressively degrade under repeated cascades. Importantly, the smaller $\sigma$ values in the upper extremes (captured in the upper whiskers) of the low-cascade regime demonstrate that LCO strongly suppresses boundary fluctuations. As LCO breaks down, both median and upper $\sigma$ values increase in the Segregated samples, reflecting more pronounced boundary structural distortions that can drive for migration. By 80 cascades, the $\sigma$ distributions converge, coinciding with the onset of grain boundary migration (Fig. 4) and a marked reduction in chemical order (Fig. 2).

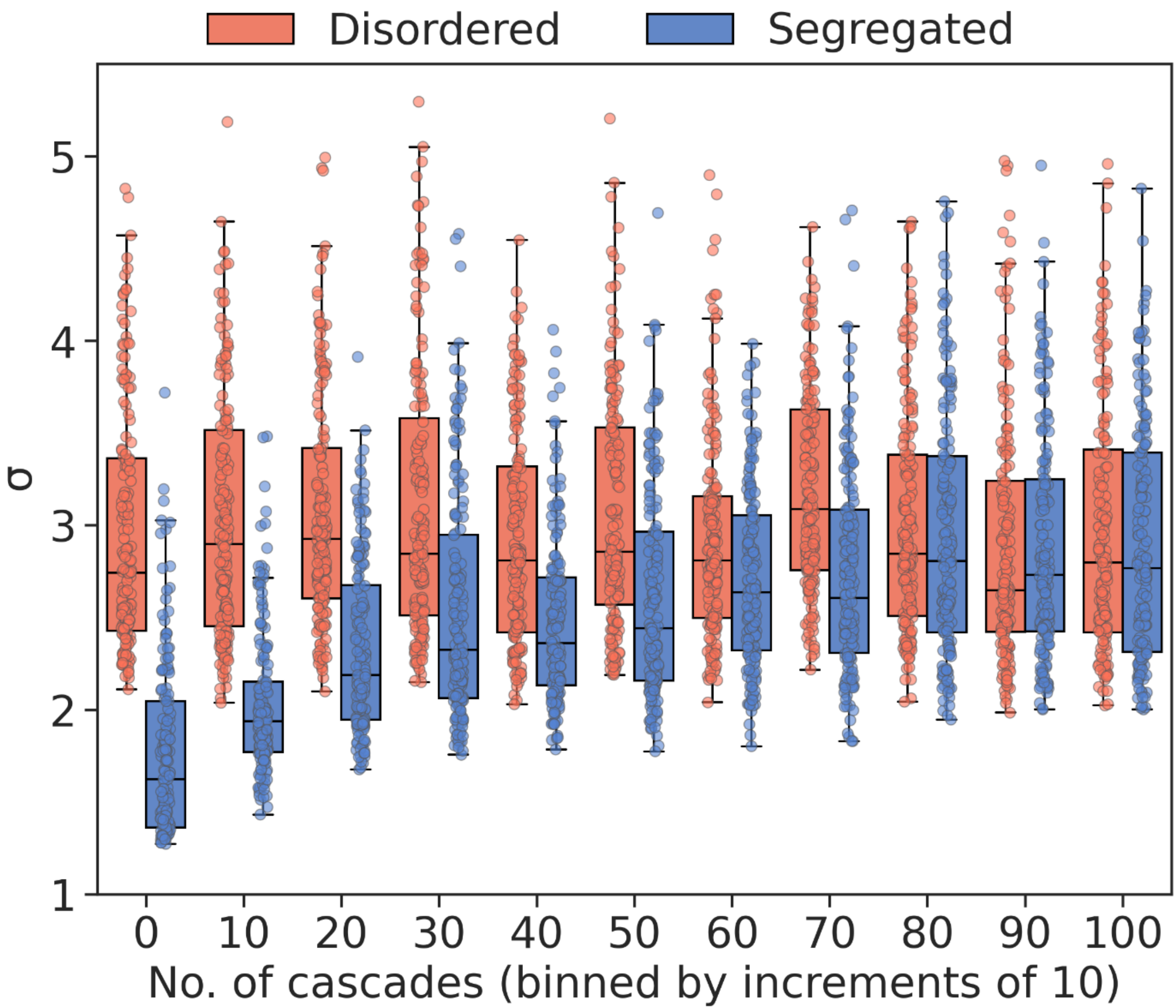

**FIG 7. Distribution of σ values for all grain boundaries after each successive cascade, binned in increments of 10 cascades and colored by the initial ordering condition of the sample.**

These results collectively show that initial suppression of grain boundary migration arises from dense, ordered regions of LCO adjacent to the interface. These regions reduce defect mobility and limit the structural fluctuations that precede boundary motion during irradiation. The magnitude of this effect is expected to vary with temperature primarily through changes in defect kinetics rather than by altering underlying mechanisms. For example, Zhang et al.[33] studied irradiated CrCoNi samples equilibrated at 1300 K and subsequently quenched to 300 K, reporting enhanced defect recombination even at lower temperature despite reduced LCO. Similar behavior at 300 K has also been observed in samples with varying degrees of LCO.[34] Because these processes primarily depend on the local energetic landscape introduced by LCO, temperature mainly influences the rates of defect evolution. Reactor materials operate across a broad temperature range depending on the system design, and the observation of significant migration suppression even at elevated temperatures suggests that the stabilization mechanism is robust. At lower temperatures, slower disordering of LCO would likely extend the duration over which ordered regions stabilize the interface.

Because the formation of these near-boundary ordered regions is closely tied to interfacial chemical state and local structural gradients,[45] the LCO-mediated suppression is also expected to vary with boundary character, as both factors depend strongly on the interface structure. For example, previous studies of binary alloys[79,80] and CCAs[81] have correlated segregation strength with misorientation angle. However, recent atomistic studies by Hua et al.[82] showed that Ni enrichment varies non-monotonically with misorientation along the <110> axis in NiCoCr, highlighting the importance of local boundary structure (e.g. symmetry and structural motif) rather than misorientation angle alone. Analysis of a structurally disordered high-angle boundary in NiCoCr with reduced Ni enrichment similarly revealed weaker near-boundary concentration waves with reduced decay length than those observed at a highly symmetric $\Sigma 11$ boundary.[45] Together, these observations suggest that both global grain boundary character and local structural features influence stabilization behavior; however, in NiCoCr the strength of near-boundary LCO appears most strongly linked to Ni enrichment. Nevertheless, LCO also develops within grain interiors independent of the interface and should still influence boundary evolution even when interfacial enrichment is limited, although the resulting effect is expected be weaker than in strongly segregating boundaries.

### 3.3. Post-irradiation effects on grain boundary structure and LCO stability

Both consecutive cascades and single PKA simulations show that grain boundaries can undergo substantial structural changes under irradiation, driven by interactions with point defects and defect clusters. Frolov et al.[83] demonstrated similar behavior in a Σ5 (310) grain boundary in pure Cu, where interstitials induce a transformation from normal kite to split kite structural units, while vacancies reverse it. Under irradiation, this process becomes highly dynamic, as interstitials and vacancies are continuously absorbed and emitted at the interface, producing heterogeneous structures. Such structural modifications can strongly influence solute segregation behavior,[25,84–86] making it critical to understand how grain boundary evolution under irradiation impacts segregation in boundary and near-boundary regions (e.g., the concentration waves observed in Fig. 1(c)). Because MD is limited to short timescales and cannot capture the diffusive processes necessary for chemical equilibration, MC/MD simulations are often used to probe long-timescale relaxation following irradiation.[46,87] Accordingly, Segregated grain boundary configurations subjected to varying levels of radiation damage were subsequently isothermally annealed at 300 K to capture structural changes and then re-equilibrated with MC/MD to assess segregation behavior.

In Fig. 8, interfacial configurations after 20, 70, and 400 cascades (post-annealing at 300 K) are compared to the initial defect-free state. Before irradiation, the boundary is dominated by normal kite motifs (red lines in Fig. 8(a)). After only 20 cascades, split kite units (blue lines in Fig. 8(b)) appear in regions where cascade damage intersects the interface. These structural transformations are likely driven by interstitial absorption, since grain boundaries act as efficient sinks for these defects, and follow the interstitial-loading mechanism described by previous studies.[83,88,89] These studies show that normal and split kite configurations represent distinct grain boundary states associated with different atomic densities, where interstitial absorption stabilizes higher-density structures and enables coexistence of multiple interfacial phases. Although both motifs persist across all damage levels, the boundary progressively shifts toward split kite dominance with increasing cascades, indicating a defect-mediated transition between competing interfacial states rather than general structural disordering. Jin et al.[18] previously showed that such interstitial loading enhances grain boundary mobility in a Σ5 (210) boundary in Cu. However, significant migration occurs only after ~80-100 cascades in the present study, while the structural transition to split kites is evident earlier. For example, by 70 cascades, split kite units dominate the structure. This indicates that the transformation itself is not the primary driver of enhanced mobility, supporting instead the role of LCO in moderating grain boundary migration.

The transition continues through 400 cascades, with normal kites becoming increasingly rare and structurally disordered regions emerging where no clear motif is identifiable. This evolution is governed by competing defect interactions under irradiation; interstitial absorption promotes split kite formation, while vacancy absorption and interstitial emission counteract the transition, leading to reversals or disordered states. Such dynamics contribute to the self-healing behavior of irradiated grain boundaries, as an interstitial-loaded boundary can emit interstitials and annihilate immobile bulk vacancies.[5] The persistence of the split kite structure at higher damage levels, despite its higher ground-state energy relative to normal kites,[83] indicates that the low annealing temperature constrains interstitial emission and bulk vacancy migration. Reversion to the normal kite structure would require removal of absorbed interstitials and long-range defect diffusion away from the interface, both of which are processes that become kinetically limited at lower temperatures. It is therefore likely that the irradiated structure is effectively frozen in a metastable state. At higher temperatures, additional relaxation through interstitial emission or Frenkel pair recombination would be expected, but under the present conditions the boundary structure remains effectively locked. We note that additional long-term annealing simulations at 1100 K did not result in widespread transformation to a split kite motif, demonstrating that high temperature alone is not enough to drive such a transition.

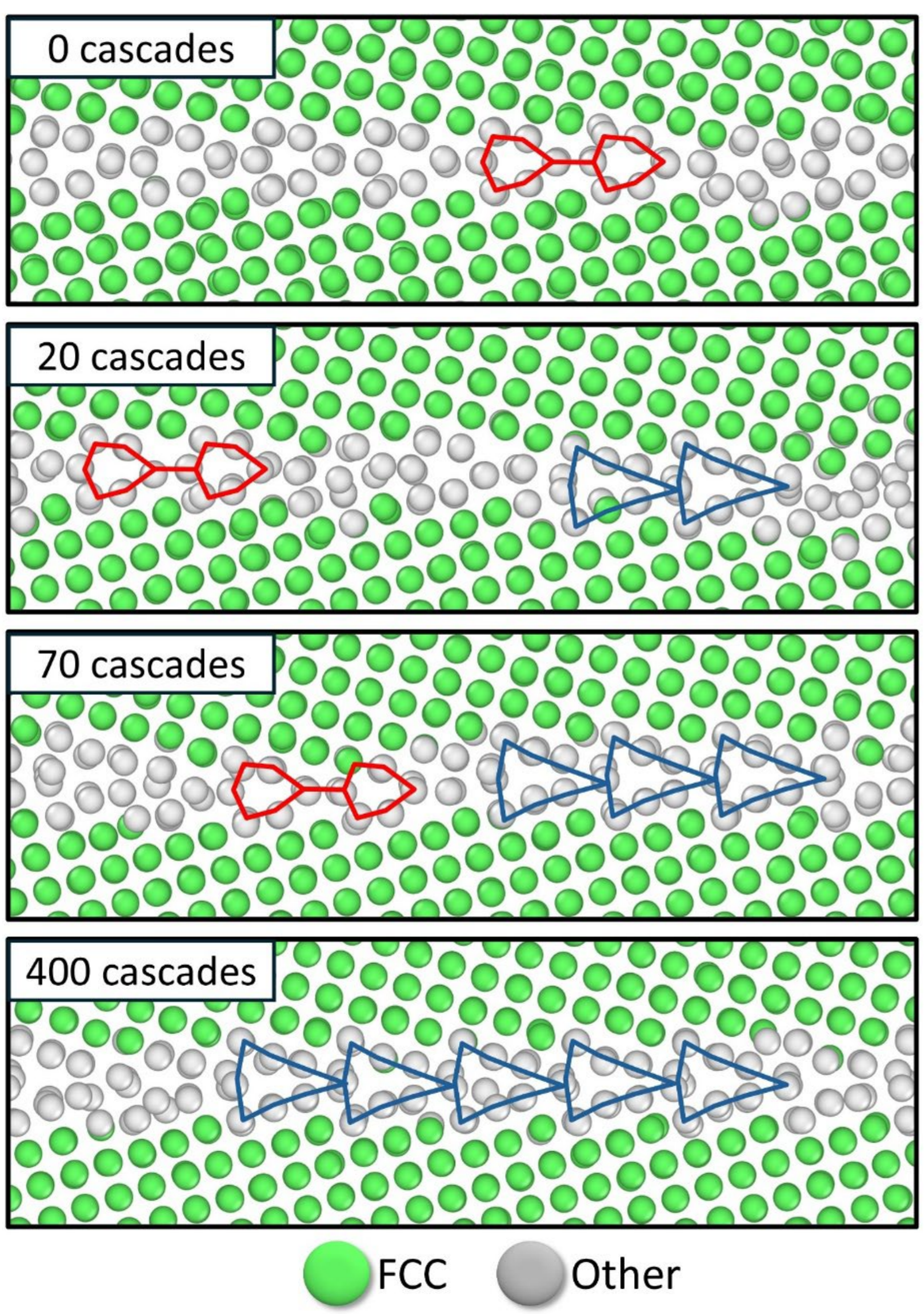

**FIG 8. Atomic snapshots of grain boundary structures prior to irradiation and after 20, 70, and 400 cascades for the Segregated sample, after annealing at 300 K and re-equilibrating with MC/MD. Normal kite and split kite structural motifs are highlighted by red and blue lines, respectively.**

Fig. 9 shows interfacial solute distributions after re-equilibration, viewed normal to the grain boundary plane. The overall composition at each damage level remains close to the initial defect-free state (i.e., Ni ~ 37.8 at.%, Co ~ 33.7 at.%, Cr ~ 28.5 at.%), though Ni segregation decreases slightly while Cr segregation increases. In contrast, the spatial segregation pattern undergoes substantial changes. Initially, the boundary exhibits well-defined linear segregation patterns with Ni clustering (green lines in Fig. 9) and Co-Cr co-segregation. Similar periodic segregation patterns, with distinct wavelengths and amplitudes, have also been observed experimentally in FeMnNiCoCr CCAs, where Ni and Mn co-segregate to some boundary regions and Cr enriches in complementary regions depleted Ni and Mn.[90] In the CrCoNi grain boundary shown in Fig. 9, the patterns observed at 0 cascades become disrupted at moderate damage levels due to irradiation-induced structural reorganization. This is particularly evident in regions where cascade-induced mixing overlaps the interface, as highlighted by the purple dashed circle after 20 cascades in Fig. 9. After 70 and 400 cascades, the original linear pattern disappears almost entirely, replaced by scattered solute distribution favoring Cr-Co and Ni-Ni ordering but lacking clear spatial organization. This morphological evolution arises from structural disordering (i.e., the transition from regular to split-kite units) induced by point defect interactions with the boundary.

Although the overall composition remains largely unaffected, irradiation-driven changes in boundary structure, such as altered free volume, local coordination, and bonding states, shift site preferences and segregation tendencies. For example, the reduction of excess free volume during a grain boundary structural transition in NbMoTaW was shown to enhance Mo and W segregation despite a strong tendency for Nb enrichment.[25] Similarly, differences in radiation-induced segregation between coherent and incoherent Σ3 boundaries in austenitic stainless steels were linked to variations in atomic structure and vacancy formation energy.[86] The extent of these changes in the grain boundary composition and structure will inevitably alter the interfacial properties, and influence the microstructure stability.

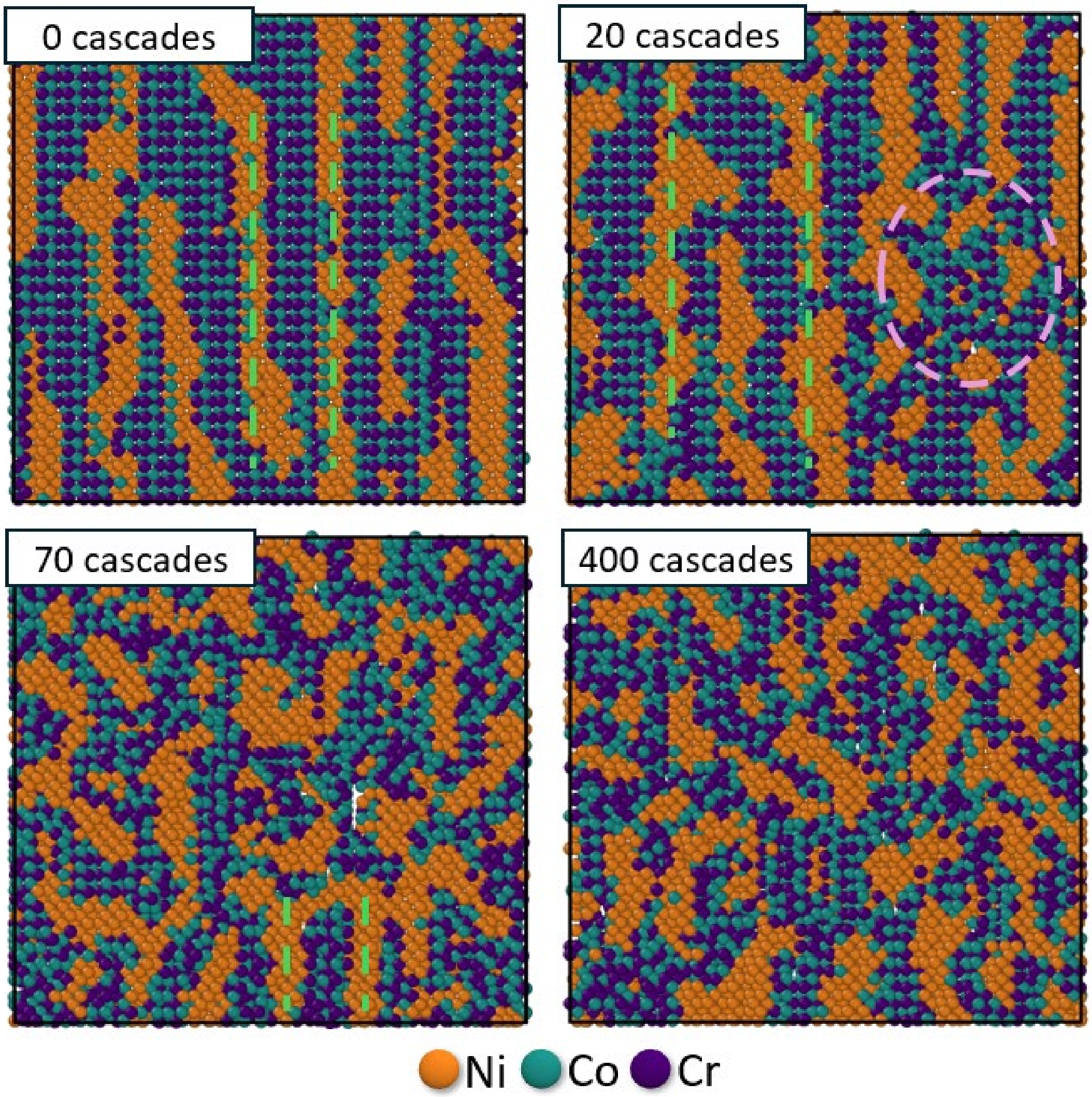

**FIG 9. Snapshots of atomic configurations after increasing damage doses in the Segregated sample, after annealing and re-equilibrating with MC/MD, viewed normal to the grain boundary plane and colored according to atom type. Green dashed lines indicate linear segregation patterns, while the purple circle shows a spatially scattered compositional pattern due to structural reorganization.**

A comparable disruption occurs in the near-boundary compositional fluctuations (Fig. 10). Compositional profiles parallel to the interface after equilibration at each damage level are shown, with black lines marking the original boundary plane location. These fluctuations are strongly coupled to interfacial chemistry,[45] with the segregation state at the grain boundary acting as a nucleation template for compositional undulations within the grains. Similar concentration waves were predicted to form at asymmetric Σ3 boundaries in body-centered cubic (BCC) NbMoTaW, suggesting that interface-driven LCO stability may be a ubiquitous feature in CCAs with strong mixing tendencies.[45] Relative to the initial state (Fig. 10(a)), the fluctuations decrease in magnitude after irradiation due to progressive chemical disordering caused by consecutive collision cascades rather than changes in bulk composition, which remains conserved throughout the additional simulations. At lower damage levels, (e.g., after 20 cascades, Fig. 10(b)), concentration waves remain visible and retain a comparable amplitude. However, by 70 cascades (Fig. 10(c)) the fluctuations are substantially diminished, and after 400 cascades, dashed lines mark new boundary positions where compositional waves re-emerge at the migrated interface. The progressive reduction in compositional waves highlights the influence of both grain boundary structure and grain boundary chemical ordering on the persistence of ordering oscillations. As the boundary becomes increasingly disordered, the local atomic environment grows more heterogeneous. Although LCO is partially recovered during equilibration, variations in chemical and structural motifs make the boundary a less effective anchor for segregation-driven concentration waves. It should also be noted that, once the boundary loses planarity (i.e., becomes wavy) after many cascades, the compositional measurements also become less reliable, though this effect is secondary to the structural changes.

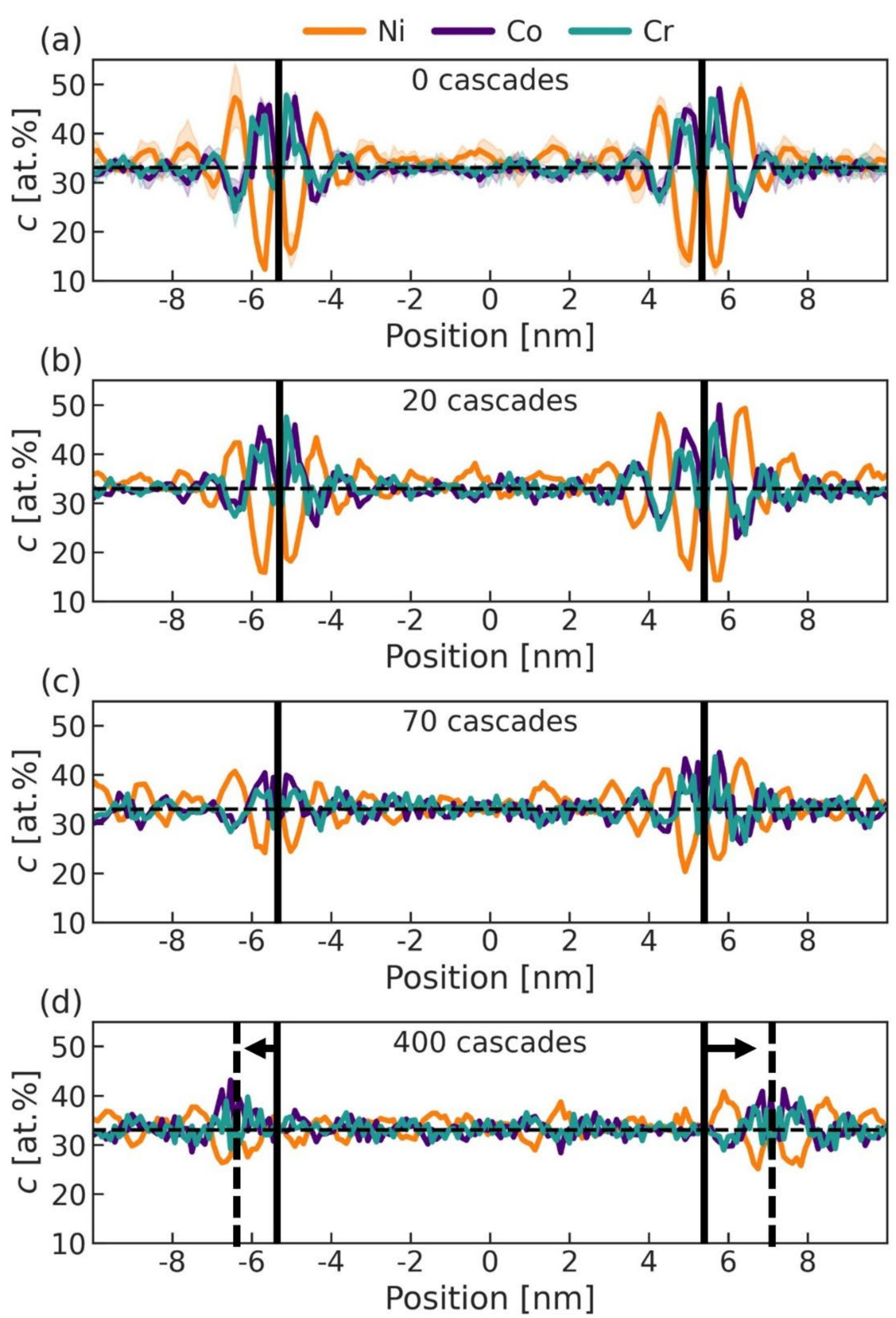

**FIG 10. Grain boundary compositional profiles following re-equilibration at 300 K after varying damage doses. Solid black lines in (a-d) show the initial grain boundary position, prior to irradiation. Grain boundary migration is illustrated with black arrows in (d), with the new grain boundary position indicated by black dashed lines.**

It is evident that the effective irradiation dose critically influences grain boundary segregation through defect absorption and interfacial disordering. The extent of this absorption depends on the interface's sink efficiency, which varies with grain boundary character.[91,92] A more comprehensive understanding, incorporating data from diverse boundary types and polycrystalline microstructures, will likely reveal a broad landscape of segregation behavior in irradiated CCAs where the coupling between dynamic re-structuring and chemical relaxation is amplified by their intrinsic chemical homogeneity. These results emphasize that the evolution of the microscopic atomic arrangements, rather than solely the behavior of larger defect clusters, represents an important and tunable factor linking atomic-scale defect processes to macroscopic material performance.

## 4. Conclusions

This study investigated the effects of LCO on the structure and stability of a Σ5 (310) symmetric tilt grain boundary in CrCoNi, leading to the following conclusions:

1) Irradiation drives dynamic changes in the spatial correlations of atomic species. In initially Disordered samples without chemical order, successive cascades enhance LCO, while in initially Segregated samples, LCO is broken down. At high doses, the magnitude of LCO for both initial conditions converge, highlighting the interplay between ballistic atomic disordering and the thermodynamic drive for re-ordering.
2) Grain boundary migration is heavily restricted in the presence of strong LCO and segregation waves. While Disordered specimens exhibit immediate fluctuations and displacement, Segregated boundaries remain immobile to moderate damage levels, after which they slowly begin to deviate from their initial positions.

3) Single PKA experiments reveal that cascade evolution differs significantly depending on the initial ordering state. The greater number of excess vacancies generated near the Disordered boundary leads to greater fluctuations and boundary curvature, which drives interfacial rearrangement. These vacancies are largely annihilated by recombination in the Segregated samples due to the reduced mobility of point defects in the cascade core.

4) Chemical equilibration on the defective states at various damage levels reveals that structural transitions and disorder alter the LCO stability, a factor that should be considered when employing materials with intrinsic chemical ordering in nuclear environments.

Overall, this study provides fundamental insight into the role of LCO in stabilizing interfaces against irradiation and into its evolution, both locally under ballistic conditions and globally during re-equilibration. As LCO is common in both FCC and BCC CCAs, this stabilization mechanism may be operative across a broad class of alloys. Our results further suggest that nanocrystalline CCAs already resistant to irradiation damage due to their high grain boundary density could gain an additional buffer against radiation-induced coarsening in systems with strong LCO. This resistance may be further strengthened by tailoring environmental conditions to optimize flux and fluence, thereby promoting LCO recovery between successive ballistic events.

**Acknowledgments**

This research was primarily supported by the National Science Foundation Materials Research Science and Engineering Center program through the UC Irvine Center for Complex and Active Materials (DMR-2011967).

# Supplementary Material

## Supplementary Note 1

Given the elevated simulation temperature (1100 K), a natural question is whether the the grain boundary migration observed in Disordered boundaries arises solely from thermal annealing. To test this, three additional Disordered bicrystals (six grain boundaries total) were annealed for an extended period without displacement cascades. Fig. S1 compares Disordered boundaries under cascades (Fig. S1(a), reproduced from the main text) with those subjected only to annealing (Fig. S1(b)). Each cascade relaxation lasted ~36 ps (with minor variation from the adaptive timestep), providing a direct temporal comparison to the annealed samples. The annealed boundaries show only slight movement from their initial position over the long simulation, whereas irradiated boundaries fluctuate substantially. This demonstrates that the migration observe under irradiation is not purely a thermal effect, though the high temperature likely enhances defect mobility within thermal spikes. Consequently, such boundary migration may not occur at lower temperatures, even under similar PKA bombardment.

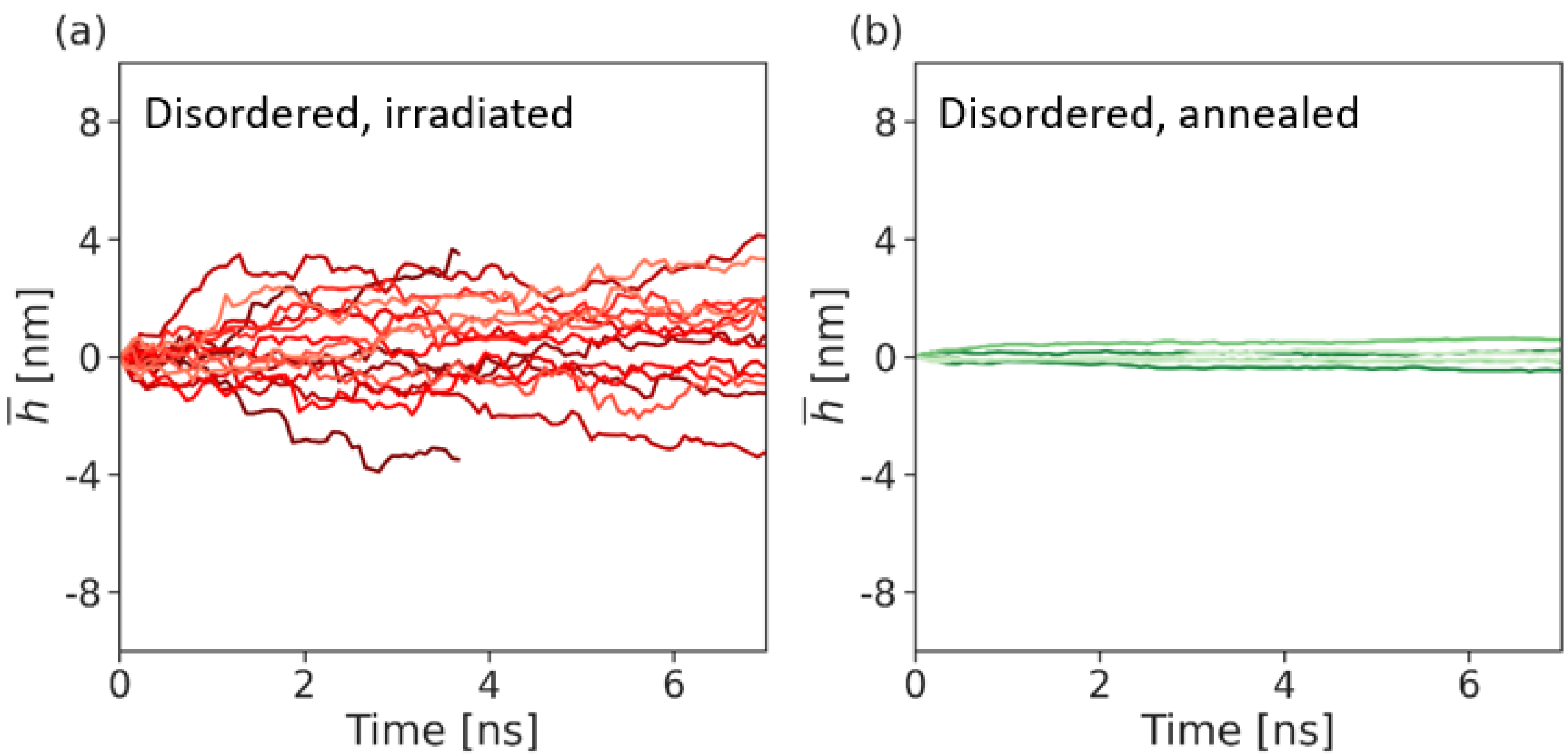

**FIG S1. Average grain boundary displacement, $\bar{h}$, for each unique simulation in Disordered (a) irradiated and (b) annealed boundary sets. Line color denotes the specific grain boundary in each bicrystal.**

## Supplementary Note 2

Additional single-PKA simulations were performed to confirm the generality of the main text results. For each boundary, a randomly selected PKA was placed 5 Å from the interface and initiated with 5 keV of energy, directed normal to the boundary. As shown in Fig. S2, a large displacement cascade develops below each boundary within 2 ps, with distinct cascade morphologies stemming from the random nature of sub-cascade formation. By 20 ps, the defect region remains extensive, containing numerous vacancies and interstitials that produce pronounced waviness in the boundary by 30 ps. This behavior arises from atomic redistribution near the interface, as many interstitials migrate to the boundary while vacancies accumulate into a cavity. In contrast, cascades in Segregated boundaries (Fig. S3) are, on average, smaller than in Disordered samples, as reflected by the mean boundary displacement, $\Delta\bar{h}$, and the standard deviation of atomic positions, σ. Snapshots at 20 ps and 30 ps show that the cascade region in Segregated samples rapidly contracts, leaving fewer point defects and producing only a minor disruption of the interface structure.

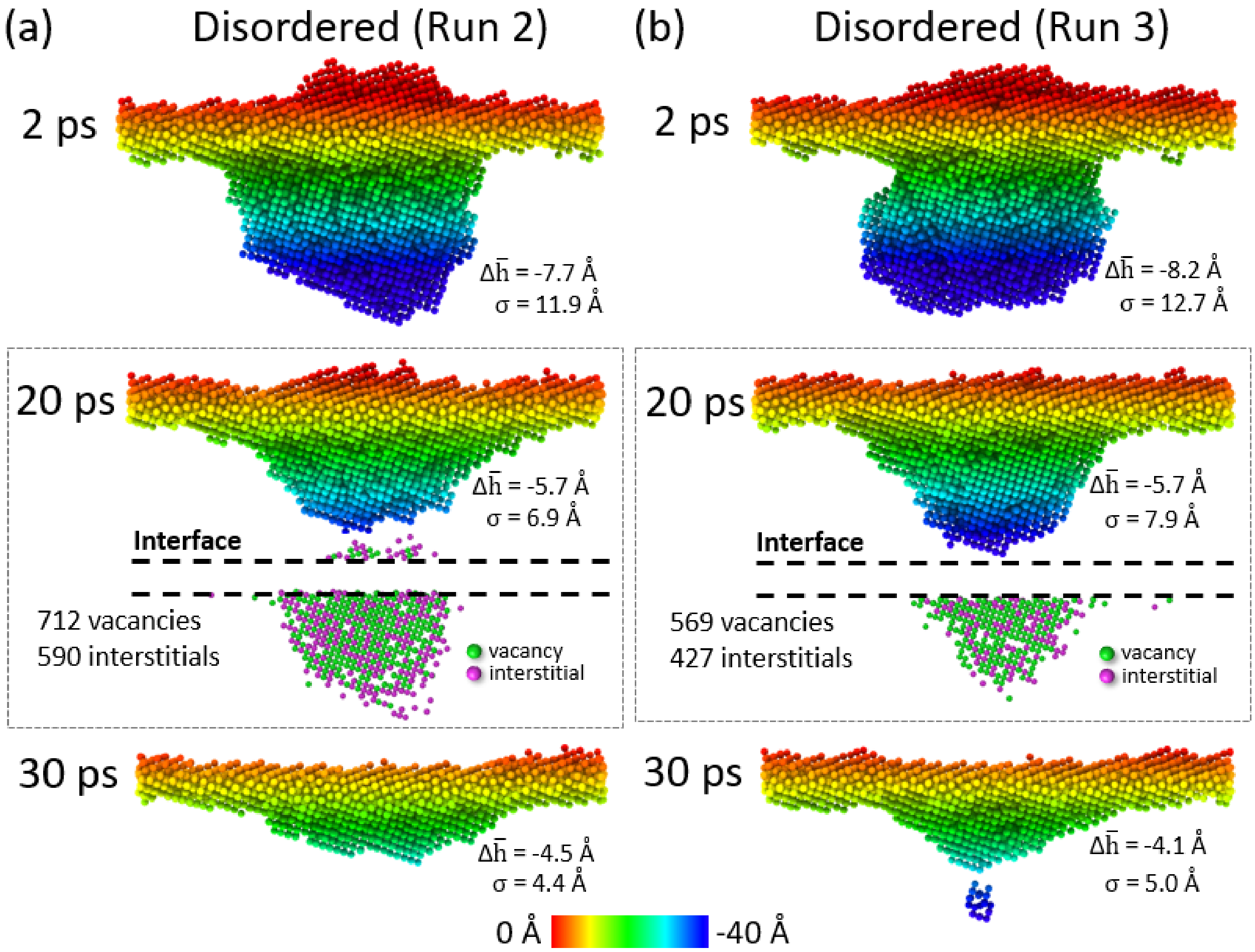

**FIG S2. Evolution of single-PKA collision cascades during the thermal spike phase for two additional Disordered boundaries. Non-FCC atoms are shown and colored by depth after 2 ps, 20 ps, and 30 ps. At 20 ps, Wigner-Seitz cell analysis highlights point defect distributions and vacancy gradients. The mean boundary displacement from the initial position, $\Delta\bar{h}$, and interfacial fluctuations, σ, are reported at each timestep to quantify migration and fluctuation behavior.**

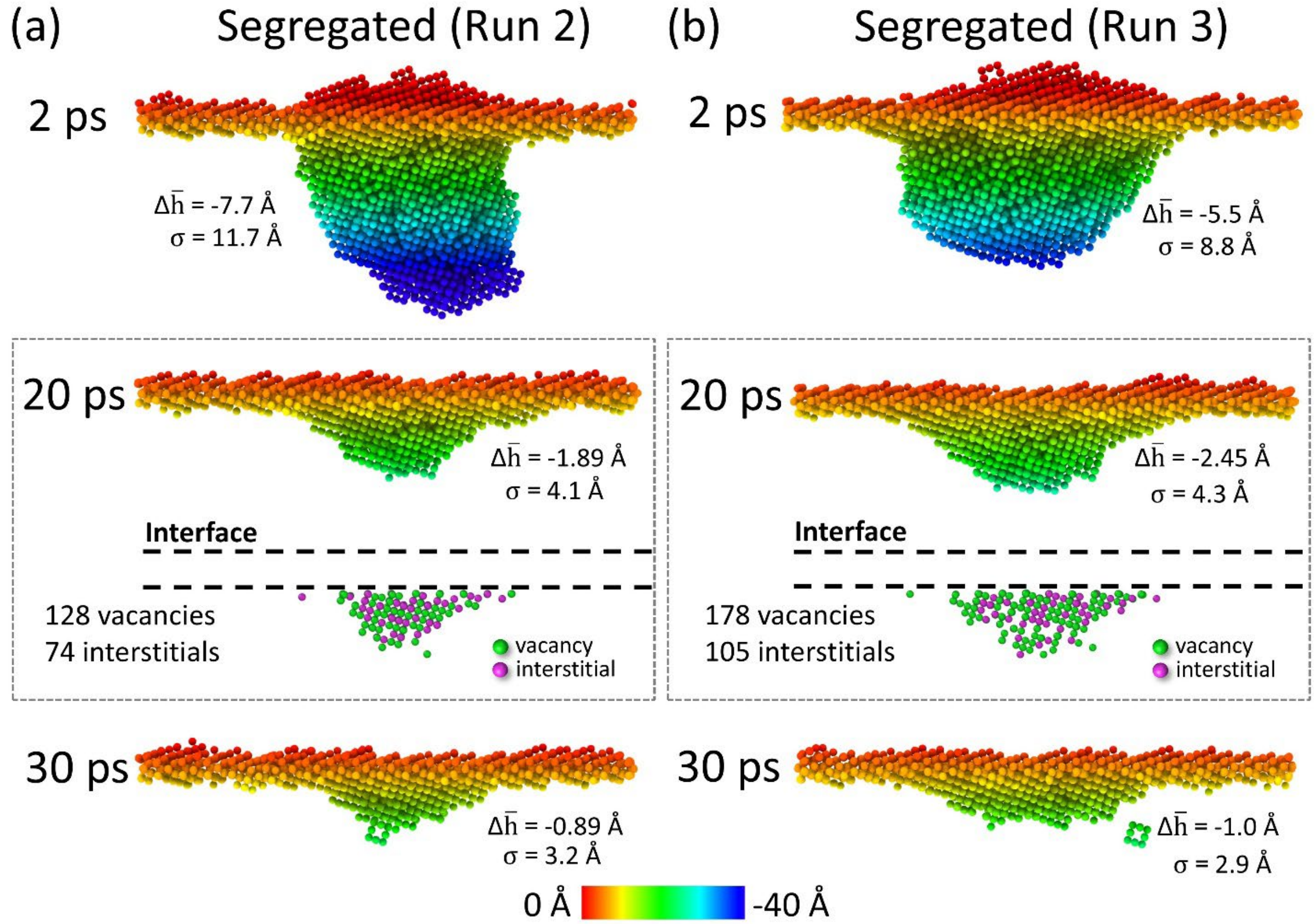

**FIG S3. Evolution of single-PKA collision cascades during the thermal spike phase for two additional Segregated boundaries.**